\documentclass[english,twoside,a4paper,10pt]{article}

\usepackage[latin1]{inputenc}
\usepackage[T1]{fontenc}
\usepackage[english]{babel}
\usepackage{amsmath}
\usepackage{amsfonts}
\usepackage{graphicx}
\usepackage{subfigure}
\usepackage{a4wide}
\usepackage{amssymb}
\usepackage{fancyhdr}
\usepackage{mathrsfs}
\usepackage{color}
\usepackage[toc,page]{appendix}

\def\ba{\begin{eqnarray}}
\def\ea{\end{eqnarray}}

\def\be{\begin{equation}}
\def\ee{\end{equation}}

\begin{document}

\title{\bf{Some remarks on  non-singular spherically symmetric space-times   }}

\author{ 
Lorenzo Sebastiani\footnote{E-mail address: lorenzo.sebastiani@unitn.it},\,\,\,
  Sergio Zerbini\footnote{E-mail address: sergio.zerbini@unitn.it}\\
\\
\begin{small}
Dipartimento di Fisica, Universit\`a di Trento, Via Sommarive 14, 38123 Povo (TN), Italy
\end{small}\\
\begin{small}
\end{small}
}

\date{}

\maketitle

\abstract{A short review on spherically symmetric static regular black holes and spherically symmetric non singular cosmological space-time is presented. Several models of regular black holes, including new ones, are considered. First, a large class of regular black holes having an inner de Sitter core  with the related issue of Cauchy horizon is investigated. Then,  black bounce space-times, where the Cauchy horizon and therefore the related instabilities are absent, are  discussed as valid alternatives of regular black holes with inner de Sitter core. Friedman-Lema\^itre-Robertson-Walker space-times admitting regular bounce solutions are also  discussed. In the general analysis concerning the presence or absence of singularities in the equations of motion, the role of a theorem due to Osgood is stressed.

}

\section{Introduction}

As well known, recently fundamental achievements in gravity research have been obtained: the  first detection of gravitational waves from binary systems of black holes (BHs) and the ``multimessanger'' signals obtained from the first observation of the collision of two relativistic neutron stars \cite{LIGO1,LIGO3,LIGO3,LIGO4}.
Furthermore,  the images of BH shadows by the Event Horizon Telescope have been reproduced (see Ref. \cite{EHT} with references therein), with further important information about the nature of these astrophysical compact objects. Other important analysis on ultra-compact stars can be found in Refs. \cite{grav,grav2, grav22, bs}.  

   The main result of these studies is that the Kerr nature of these compact objects is confirmed within a small uncertainty. Since Kerr BH solution   is a mathematical   vacuum solution of General Relativity (GR), one may  conclude that it might be a  good but approximate  description of reality, and the possible existence of other compact objects  with horizon may  not be ruled out. We should also stress that many conceptual problems still exist about the nature of Kerr BHs, namely the central singularity problem and so on. In fact, Kerr BH is the unique vacuum solution of GR and admits  a non physical singularity. This is a consequence of the GR singularity theorems.

Thus, one needs alternative ``stuffs'' to avoid the problem: for example, regular black holes associated with
  exotic sources in order to bypass GR singularity theorems. We also should note that regular BHs exist in vacuum (absence of matter) if one goes beyond GR in a modified gravity framework.

  With regard to this issue, maybe it is necessary to clarify that in our paper non-singular space-times are space-times free of curvature divergences. We would remind that in GR, the singularity theorems state that the presence of geodesic incompleteness is associated with the appearance of physical singularities.  For us, the presence of severe pathologies in the curvature invariants are sufficient condition to regard the space-times as singular, being aware that  geodesic incompleteness and curvature singularities might be, in some cases, not equivalent concepts

  In our paper we will discuss several aspects of regular black holes,
  namely BHs where the central singularity is absent, due to the presence of a de Sitter (dS) core or the presence of a fundamental length. 
  
  With regard to this, we will first analyze a class of solutions with delta-regularized source. These solutions are interesting, since they present a Schwarzschild-like behaviour at large distances and for suitable choices of parameters the horizons (event and Cauchy horizons) may be absent. The related  metrics are asymptotically flat. In the BH case, However, the regular behaviour of the metric is not a sufficient condition to avoid singularities in the curvature invariants and their derivatives. Thus, we will discuss the Sakarov criterion, which allows us to identify a restricted class of regular BH solutions, for which the scalar invariants and their covariant derivatives are every where bounded. 
  
  Furthermore, in general in the BH case, the presence of a dS core brings also to the existence of an inner Cauchy horizon which may lead to some instability. In this respect, a class of alternative BHs is given by Black Bounce space-times,  metrics which are asymptotically flat and the Cauchy horizon is absent. For these metrics,  the central singularity is absent thanks to the introduction of a minimal length scale in the metric components.  
  
  We will also show how the various BHs can be recovered in the framework of regularized Lovelock Lagrangians, a class of Lagrangians that have been object of recent interest. 
  
  Finally, we will pose attention on non-singular cosmological models, where we will investigate the absence of singularities in cosmology with some  general considerations on the proprieties of field's equations of the theories under investigation.  

 The paper is organized as follows. In Section {\bf 2} we will revisit the formalism of Spherically Symmetric (SS) space-time and we will focus our attention on static black hole and wormhole (WH) solutions. In Section {\bf 3} we will discuss a special class of black holes with inner de Sitter core in the presence of delta-like regularized source and some applications of covariant Sakarov Criterion. In Section {\bf 4} alternative black holes described by Black Bounce space-time are analyzed as valid alternative of non-singular black holes without Cauchy horizon. In Section {\bf 5} black holes in four dimensional regularized Lovelock Lagrangians are discussed, while Section {\bf 6} is devoted to cosmological models free of space-time singularity. Conclusions and final remarks are given in Section {\bf 7} and in Appendix additional material on the Painlev\`e gauge and its application to Hawking radiation regarding generic static BHs and WHs is reported. 

In our convention, the speed of light $c = 1$ and the Newton Constant $G_N=1$, such that the Planck mass is $M_{Pl}^2=1/8\pi$. We also adopt the "mostly plus" metric convention.

\section{ Black Holes and Wormholes in Static Spherical Symmetric space-time}

Here we recall the Kodama-Hayward invariant formalism (see for example \cite{kodama,Sean89,noi,vanzo}) 
 for a generic four dimensional Spherical Symmetric Space-time (SSS). The related metric reads
\begin{equation}
ds^2=\gamma_{ab}dx^a dx^b+ r(x^a)^2 dS^2\,,
\end{equation}
where $dS^2$ is the metric of a two-dimensional sphere, 
$\gamma_{ab}$ is the metric tensor of the two dimensional space-time (the normal metric)  with coordinates $x^a\,, a=0,1$, and $r\equiv r(x^a)$ is the areal radius and is a  scalar quantity which depends on the coordinates of the normal space-time. Another relevant scalar, related to the variation of a surface with radius $r$,     is given by
\begin{equation}
 \chi(x^a)=\gamma^{a b}\partial_a r\partial_b  r\,,\label{chi0}
\end{equation}
which defines a (dynamical)  trapping horizon by
\begin{equation}
\chi(x_H)=0\,,\quad \partial_a\chi(x_H)>0\,,\quad a=0,1\,. \label{Hcond}
\end{equation}
The second condition is required in order to preserve the metric signature out of the horizon.
A related scalar quantity is the  Hayward surface gravity, 
\begin{equation}
\kappa_H= \frac{1}{2\sqrt{-\gamma}} \partial_a\left(\sqrt{-\gamma}\gamma^{ab} \partial_b r\right)_H \,,\label{kHay}
\end{equation}
where $\gamma$ is the determinant of the two-metric $\gamma_{a b}$ and the pedex $`H'$ denotes a quantity evaluated on the horizon.

Finally,   the   Kodama vector is 
\begin{equation}
K^a=\frac{\epsilon^{a b}}{\sqrt{-\gamma}}\partial_ar \,,\label{K}
\end{equation}
where $\epsilon^{ab}$ is the volume-form associated with the two-metric $\gamma_{a b}$.
The Kodama vector is orthogonal to the normal space-time and is covariant conserved in a generic SS space-time, namely
\begin{equation}
\nabla_\mu K^\mu=0\,.
\end{equation}
Note that $\sqrt{-\gamma}  $ has to be well defined such that $-\gamma >0$.

  \subsection{ The static case}

The Static SSS case  is  well understood and investigated. In the    Schwarzschild gauge the metric reads
\begin{equation}
ds^2=-A(r)dt^2+\frac{dr^2}{B(r)}+ r^2 dS^2\,,
\label{sg00}
\end{equation}
where $A\equiv A(r)$ and $B\equiv B(r)$ are metric functions of the radial coordinate $r$ only.
In what follows, 
we denote with the prime index the derivative with respect to $r$. 

Note that the determinant of the metric $g$ is given by $\sqrt{-g}=\sqrt{\frac{A(r)}{B(r)}}r^2$, such that $\frac{A(r)}{B(r)}>0$. The  invariant quantity $\chi\equiv \chi(x^a)$ and the Kodama vector are well defined and read
\begin{equation}
\chi=B(r)\,, \quad K^\mu=\left( \sqrt{\frac{B(r)}{A(r)}},0,0,0 \right)\,.
\end{equation}
The Kodama energy $\omega$  associated with a  test particle with four-momentum  $p_\mu=\partial_\mu I$, $I$ being the action, is given by 
\begin{equation}
\omega=-K^\mu p_\mu= \sqrt{\frac{B}{A}} E\,,
\label{k20}
\end{equation}
where   $E=-\partial_t I$ is the test particle Killing energy.

 Black holes and wormholes  possess a trapping (event) horizon with $ \chi_H= 0$. If $A(r_H)=B(r_H)=0 $ one has a BH. If  $B(r_H)=0$, but $A(r_H) \neq 0 $, one has to deal with a WH \cite{Morris,visser, visser2, visser3, visser4, visser5, visser6, SeanWH}. If $\chi$ is  never vanishing, one obtains an  horizonless compact object (HCO) .

The Kodama vector  may provide an invariant way to distinguish between BHs and WHs \cite{lor}. 
In the Schwarzschild  static gauge the trapping horizon is defined by
\begin{equation}
B(r_H)=0 \,.
\label{sths}
\end{equation}
Thus,
we may  define a  static black hole as a SSS solution  where the Kodama energy  
in (\ref{k20})
evaluated on the horizon is not vanishing (see also the Appendix for a derivation of the Hawking temperature). This is the case of a black hole where $A(r)$ is  proportional to $B(r)$ and $A(r_H)=0\,,A'(r_H) >0$ (event horizon). 

The  Hayward surface gravity  associated with a trapping horizon is
\begin{equation}
\kappa_H=\frac{B'(r_H)}{2}\,.
\end{equation}
In the static case one can define a time-like Killing vector field 
with an associated Killing surface gravity, 
\begin{equation}
\kappa_H^{K}=\frac{\sqrt{B'(r_H) A'(r_H)}}{2}\,.
\label{k2}
\end{equation}
In general, the Killing surface gravity differs from the Hayward surface gravity due to a different renormalization of the Killing vector with respect to the Kodama vector. However,
for most of known BHs, $A(r)=B(r)$, and they coincide.

We recall that if $A(r)\neq 0$,  in particular  on the horizon, one has a 
 static regular wormhole a la Morris-Thorne.  Thus, $\omega_H$  is  vanishing.
The Hayward surface gravity 
\begin{equation}
\kappa_H=\frac{B'(r_H)} {4}\,, 
\label{k222}
\end{equation}
  implies a minimum value $r_H$ for $r$,  throat or mouth of a  traversable wormhole ($\kappa_H>0$), such that $r>r_H$. 
Moreover, if $A(r)$ is vanishing at some point, one has  singular WHs. A well known example is the Brans-Dicke WH \cite{BS, BS2, BS3, BS4, BS5, BS6, BS7, BS8}. 

An other interesting example is given by a variant of Damour and Solodhukin metric \cite{DS},
\begin{equation}
ds^2=-\left(1-\frac{2m}{r}\right)dt^2+\frac{dr^2}{\left(1-\frac{2m+b^2r}{r}\right)}+r^2 d S^2
\,,\label{DSZ}
\end{equation}
where $m$ is a mass parameter and  $b^2$ is a dimensionless   arbitrarily small parameter. Here,  $r_H=\frac{2m}{1-b^2}$,
but $A(r_H)=b^2 >0$ and 
 the Kodama energy  vanishes on the throat, namely $\omega_H=0$, and we are describing a  static  wormhole.
Since  $r >r_H$, the  singularity $r=0$ is not present.

Static WHs are  non singular objects, but
in General Relativity they can be obtained only in the presence of a source  of exotic matter which violates the energy conditions and they are generally unstable (see also the recent work in Ref. \cite{Casaz} where vacuum WH solutions are found in the presence of trace anomaly contribution).

\subsection{ Effective Fluid Models in General Relativity}

 The simplest way to go  beyond GR is to consider the
 Einstein's equations in  presence of  effective  relativistic anisotropic fluids.
The Einstein's equations are 
\begin{equation}
G^{\mu \nu}=8\pi T^{\mu \nu}\,,
\label{ee0}
\end{equation}
with
\begin{equation}
 T^{\mu \nu}=(\rho+p_T)u^\mu u^\nu+p_Tg_{\mu \nu}+(p_r-p_T)C^\mu C^\nu\,,
\label{ee}
\end{equation}
 where $u^\mu u_\mu=-1$ is a time-like vector and $C^\mu C_\mu=1$ is the  anisotropy space-like vector. Moreover, $\rho$ is the energy density of fluid, $p_r$ the radial pressure and $p_T$ the trasversal pressure.
Equations of motion read
\begin{equation}
r B'+B-1=-8\pi r^2 \rho\,,
\label{EOM1}
\end{equation}
\begin{equation}
\frac{A'}{A}-\frac{B'}{B}=\frac{8\pi r (\rho+p_r)  }{B}\,.\label{EOM2}
\end{equation}
The Tolman-Oppenheimer-Volkov equation leads to
\begin{equation}
 p_r'+\frac{\rho+p_r}{2}\frac{A'}{A}=
  \frac{2 (p_T-p_r  )}{r}   \,.\label{TOeq}
\end{equation}
In principle, given the energy density $\rho$,  the metric function $B(r)$ is computable.
 Chosen the fluid Equation of State (EoS) $p_r=p_r(\rho)$, the metric function  $A(r)$ is also computable.
 Finally, the Tolman-Oppenheimer-Volkov equation gives the form of $p_T$.
    Alternatively, chosen $A(r)$ and $B(r)$,  $\rho, p_r, p_T$ are computable and we can reconstruct the fluid which supports a given SSS metric.

 \section{Regular black hole solutions with inner de Sitter core}
 
 Since space-time singularities are very problematic to treat, a lot of investigations have been carried out 
 on non-singular black hole solutions, where the central singularity which characterizes the Schwarzschild metric is removed and substituted by a De Sitter core. The resulting solution is a non-vacuum solution of Einstein's equations.
 Explicit examples have been provided by Bardeen, Hayward, and many others (for a partial list see  Refs. \cite{Bardeen,AA,seanBH,Bonanno,D, Nico,ans,balakin,bronnikov,Ale,CZ,CCZ,mi,Bal,Fan,Carballo-Rubio,Co,Cadoni}, and references therein). Other related results can be found in \cite{C1,C2,C3}.
 
 Since on BH horizon the conditions $A(r) = 0$ and $B(r) = 0$ must be satisfied for the same value of $r = r_H$, a direct consequence of Eqs. (\ref{EOM1})--(\ref{EOM2}) is that $p_H = -\rho_H$. Thus, a simple way to obtain non-vacuum BH solutions is given by the generalization of this condition, namely one assumes the EoS
 $p=-\rho$ such that  $A(r)=B(r)$.
 Furthermore, it is convenient to write the metric functions in the form 
\begin{equation}
A(r)=B(r)=1-\frac{2m(r)}{r}   \,, \label{regB}
\end{equation}
where $m\equiv m(r)$ is a mass function depending on $r$. For $m(r)=m$ one gets the Schwarzschild solution.
The Ricci scalar reads
\begin{equation}
R=\frac{2}{r^2}(1-A(r))-4\frac{A'(r)}{r}-A''(r)=\frac{4m'(r)}{r^2}+2\frac{m''(r)}{r} \,,  
\end{equation}
while the other curvature invariants are
\begin{equation}
R_{\mu \nu }R^{\mu \nu }=   8\frac{m'(r)^2}{r^4}+2\frac{m''(r)^2}{r^2}\,,
\end{equation}
\begin{equation}
R_{\mu \nu \alpha \beta}R^{\mu \nu \alpha \beta }=  
48\frac{m(r)^2}{r^6}+O\left(\frac{m''(r)^2}{r^2},\frac{m'(r)}{r^2}\right) \,. 
\end{equation}
As a result, for $r \rightarrow 0$, , in order to avoid the central singularity, one has to take  $m(r)=r^3C+O(r^4)$, $m'(r)=3r^2C+O(r^3)$, $m''(r)=6rC+O(r^2)$, $C$ being a constant, and the curvature invariants are finite at $r=0$. In this way, one gets for small values of $r$,
\begin{equation}
A(r)=1-2Cr^2+...  
\end{equation}
This is the so called Sakarov Criterion:  an interior de Sitter core leads to BH with finite curvature invariants. Related to this there is the Limiting Curvature Conjecture (see Ref. \cite{Frolov} and references therein).

A very popular example is given by   Poisson-Israel-Hayward  BH \cite{seanBH}
\begin{equation}
A(r)=B(r)=1-\frac{2mr^2}{r^3+\ell^3}\label{h1}\,,
\end{equation}
and 
\begin{equation}
A(r)=B(r)=1-\frac{2mr^2}{r^3+\ell^2m}\label{h2}\,,
\end{equation}
$m$ and \,$\ell$ being  constants. The metric exhibits a de Sitter core for small values of $r$, while for large values of $r$ one recovers the behaviour of Schwarzschild space-time. In order to deal with a BH, the mass $m$ has to satisfy  the  inequality $m>m_c=3^{3/2}\ell/4$. For $m< m_c$, there is not an horizon and one has a simple example of  horizonless compact object  (HCO), very similar to the  Gravastar of Mazur-Mottola, which is characterized by a dS core with stiff matter shell (interior) and Schwarschild behaviour at large distances (exterior).

Unfortunately,  solutions with $A(r)=B(r)$ and dS core
suffer of two problems. 
  The first one is that the  square of the radial speed of sound $v_s^2$ is negative, a possible signal of  instability. This is a direct consequence of the fact that
  $p_r= -\rho $. Thus, $ v^2_s=\frac{d p_r}{d \rho}=-1 <0$. 

The second problem is related  to the presence of an inner Cauchy horizon with associated   instability related to   mass inflation \cite{Israel} and  kink instability \cite{Har}. Mass inflation is the result of the exponential growth of mass parameter of the solution due to the crossflow of infalling and outgoing radiation perturbations  and it is strictly connected with the presence of a Cauchy horizon (see the recent papers with references therein in \cite{DiFilippo, DiFilippo2}).  Moreover, regarding mass inflation issue see also Refs. \cite{Doku,Bona,Herm}, and Ref. \cite{Berti} in the case of quadratic gravity.

Very recently Visser and co-workers \cite{Car} have constructed a regular BH with a dS inner core and asymptotically flat,  working in the Schwarzschild gauge (\ref{sg00}) with metric functions $A(r)=B(r)$ such that 
\begin{equation}
A(r)=\frac{(r-r_H)(r-r_C)^3}{N(r)}\,,    
\end{equation}
with $r_C <r_H$ and $N(r)$ a suitable function such that there exists a dS core and $A(r)=1-\frac{2M}{r}+..$ for large values of $r$.
There are two horizons, the event horizon at $r=r_H$ and a Cauchy horizon at $r=r_C$, which is, however a triple zero of $A(r)$. As a consequence, the surface gravity associated with it is vanishing and the mass inflation is absent.
This result has some similarities with the absence of mass inflation in higher order derivative gravity obtained in Ref. \cite{Berti}.

\subsection{Regular black holes with delta-like regularized source}

Among the regular BHs with de Sitter core, there exists a specific class we are going to discuss. The main idea, well justified by physical considerations, is to deal with densities which are related to regularization of Dirac delta distribution in spatial dimension $D=3$. 

With regard to this issue, we recall that if we have an even integrable positive function, namely
$f(-\Vec{x})=f(\Vec{x})$,  $f(\Vec{x})>0$, and with
\begin{equation}
    \int_{R^3} d\Vec{x}f(\Vec{x})=C <\infty \,,\label{up}
\end{equation}
then, 
\begin{equation}
\lim_{\epsilon\rightarrow 0}
\frac{1}{C \varepsilon^3}f\left(\frac{\Vec{x}}{\varepsilon} \right) = \delta(\Vec{x}   )\,.
\label{epsilon0}
\end{equation}
Let us use the Schwarzschild gauge (\ref{sg00}) and take now the simple EoS
$p=-\rho$ such that  $A(r)=B(r)$ as previously discussed. We will assume the form (\ref{regB}) for our metric functions.
Moreover, we take as density
\begin{equation}
\rho=\frac{M}{C \varepsilon^3}f\left(\frac{r}{\varepsilon}\right)  \,.  
\end{equation}
Here, $f(r)$ is an even integrable positive function which satisfies (\ref{up}),
$M$ is a mass parameter, and $\varepsilon$ the regularization parameter, such that $\rho$ is a delta-like regularized energy density source. 

By using (\ref{EOM1}), 
the mass function results to be defined as 
\begin{equation}
m(r)=4\pi \int_0^r dy y^2 \rho(y)   \,.
\end{equation}
Now it is convenient to introduce a new function $g(r)$, dubbed ``g-function'', defined by $2m(r)=r^3 g(r)$. Then one has,
\begin{equation} 
 g(r)=\frac{8\pi}{r^3}   \int_0^r dy y^2 \rho(y) \,,  \quad A(r)=B(r)=1-r^2 g(r)\,.
\label{gfunc}
\end{equation}
Therefore, the g-function associated with a delta regularized density 
\begin{equation}
g(r)=\frac{8\pi }{ r^3}\int_0^r dy y^2    \frac{M}{C \varepsilon^3}f\left(\frac{y}{\varepsilon}\right) \,,
\end{equation}
namely
\begin{equation}
g(r)=\frac{8\pi M}{C r^3}\int_0^{r/\varepsilon} dy y^2 f(y) \,.
\end{equation}
It is easy to show that  one recovers  the Schwarzschild limit for large values of $r$, namely $r^2g_G=\frac{2M}{r}+ O(1/r)$. 
In order to obtain a (hyper) dS core for small $r$, one has to assume the smooth behavior for $f(r)$ at $r=0$, namely
$f(r)\simeq r^a(c_0+ c_1 r+... )\,, a>0$ and $c_{0,1,...}$  constants. Thus,
\begin{equation}
g(r)\simeq\frac{8\pi M r^a}{(3+a)C \varepsilon^{3+a}}(c_0+O(r))  \,.  \end{equation}
For $a=0$, one has a dS core. Moreover, since $A(r=0)=1$ and $ A(r\rightarrow\infty)=1$, an even number of zeros exist. For example, according to the values of parameters $\varepsilon$, $a$, and $M$,  there are two horizons or no horizons. Let us see for some examples.

As a first example, one may take the family of  the well known Gaussian delta-like generating functions $f(\Vec{x})=|\Vec{x}|^a e^{-\Vec{x}^2}$
with $a$ a non negative real number.
Then,
\begin{equation}
C_a\equiv\int_0^\infty r^{a+2}e^{-r^2} dr=  2\pi \Gamma\left(\frac{3+a}{2}\right)\,,   
\end{equation}
where $\Gamma(\alpha)$ is the Gamma function,
and the related energy density is well defined and there is a dS core,
\begin{equation}
 \rho_G=\frac{M}{C_a \varepsilon^3}\left(\frac{r}{\varepsilon} \right)^a  e^{-\frac{r^2}{\varepsilon^2}} \,.  
\end{equation}
In fact, the associated g-function reads
\begin{equation}
 g_G(r)=   \frac{2 M}{\Gamma\left(\frac{3+a}{2}\right) r^3}
 \gamma\left(\frac{3+a}{2},\frac{r^2}{\varepsilon^2}\right)\,,\label{gG}
\end{equation}
where now $\gamma(\alpha,z)$ is the (lower) incomplete Gamma function. For example, for $a=1$, one simply obtains
\begin{equation}
g_G(r)=\frac{M}{ r^3}\left(
1-\text{e}^{-\frac{r^2}{\varepsilon^2}}\left(1+\frac{r^2}{\varepsilon^2}\right)
\right)\,.
\end{equation}
We can check the presence of a dS core for small $r$ when $g_G(r)$ in (\ref{gG}) assumes the form,
\begin{equation}
g_G(r)\simeq  \frac{2 M}{\Gamma(\frac{3+a}{2})}  \frac{r^{a}}{\varepsilon^{a+3}}\left(c_o+c_1 \frac{r^2}{\varepsilon^2}+c_2+
\frac{r^4}{\varepsilon^4}+
...   \right)\,,\label{gGsmallr}
\end{equation}
with $c_{0,1,2,...}$ constants coefficients depending on $a$. In the case of $a=1$ one has 
\begin{equation}
g_G(r)\simeq  \frac{M r}{\varepsilon^4}
\,.
\end{equation}
For  $a=0$, the related BH has been discussed in Ref. \cite{Nico}. 

 Another class of regular BHs may be constructed making use of the other well known delta-generating function, the Lorentzian one, namely $f(\Vec{x})=\frac{|\Vec{x}|^a}{(x^2+1)^{N/2}}$, where $a$ is a real number and $N$ a natural number. The integrability in $D=3$ requires $N>3+a>0$, and we take $a>0$. The normalization constant results to be,
\begin{equation}
    C_N=4\pi \int_0^\infty \frac{r^{2+a}}{(r^2+1)^{N/2}}dr=2\pi \Gamma\left(\frac{3+a}{2}\right)\frac{\Gamma(\frac{N-3-a}{3})}{\Gamma(N/2)}\,.
\end{equation}
The associated energy density is
\begin{equation}
\rho_N=\frac{M}{C_N}\frac{\varepsilon^{N-3-a}}{(r^2+\varepsilon^2)^{N/2}}\,.
\end{equation}
The related g-function reads
\begin{equation}
g_N(r)=\frac{8\pi M}{C_N r^3}\int_0^{r/\varepsilon}     \frac{y^{2+a}}{(y^2+1)^{N/2}}dy= \frac{8\pi M \varepsilon^{-a-3}}{c_N (3+a)}r^a F\left(\frac{3+a}{2}, \frac{N}{2} ;\frac{5+a}{2},-\frac{r^2}{\varepsilon^2}\right) \,,
\end{equation}
where $F(\alpha,\beta;\gamma,z)$ is the Gauss hypergeometric function. For general $N>3+a$, it is easy to show that for small $r$ there is a dS core, 
\begin{equation}
    g_N(r)\simeq \frac{8\pi M \varepsilon^{-a-3}}{C_N(3+a)}r^a\left(1- \frac{(3+a)N}{2a+10}\frac{r^2}{\varepsilon^2}+O(r^4)\right)\,.
\end{equation}
On the other hand, for large values of $r$, one has $r^2 g_N(r)=\frac{2M}{r}+...$ and $M$ can be identified with the BH mass. 

Let us consider some specific values for $N$ and $a$.
We start with $a=0, N=4$ such that
\begin{equation}
g_4(r)=\frac{4M}{\pi r^3} \left(\arctan \left(\frac{r}{\varepsilon}\right)-\frac{\varepsilon r}{r^2+\varepsilon^2}  \right)  \,.  
\end{equation}
This corresponds to a BH discussed by Dymnikova \cite{D}.

Next example is $N=5, a=0$. With this choice, one has
\begin{equation}
g_5(r)=\frac{M}{(r^2+\varepsilon^2)^{3/2}}    \,.
\end{equation}
This corresponds to the well known Bardeen BH \cite{Bardeen}.

Of course, one may continue and for $N=6, a=0$, we get 
\begin{equation}
g_6(r)=\frac{4M}{\pi r^3} \left(\arctan \frac{r}{\varepsilon}-\frac{\varepsilon r(r^2-\varepsilon^2}{(r^2+\varepsilon^2)^2}   \right)  \,.  
\end{equation}
The cases $N=7,8,9,  a=0$, present no difficulties.

We conclude this subsection with other two examples. The first example is related to the following choice,
\begin{equation}
f(\Vec{x} ) =\frac{3}{(1+|\Vec{x}|^3)^2}  \,.
\end{equation}
The normalization constant is $C=4\pi$ and the g-function turns out to be
 \begin{equation}
  g(r)=  \frac{24 \pi M}{C r^3}\int_0^{r/\varepsilon}dy \frac{y^2}{(1+y^3)^2} \,.
\end{equation}
Thus,
\begin{equation}
g(r)=\frac{2M}{r^3+\varepsilon^3}    \,,
\end{equation}
which corresponds to Poisson-Israel-Hayward BH in (\ref{h1})--(\ref{h2}).

For the second example we take 
\begin{equation}
f(\Vec{x} ) =\frac{3}{(1+|\Vec{x}|)^4}  \,.
\end{equation}
The normalization constant is $C=4\pi$ and the g-function reads
\begin{equation}
  g(r)=  \frac{24 \pi M}{C r^3}\int_0^{r/\varepsilon}dy \frac{y^2}{(1+y)^4} \,.
\end{equation}
Thus,
\begin{equation}
g(r)=\frac{2M}{(r+\varepsilon)^3}    \,,
\end{equation}
associated with a regular BH investigated by Fan and Wang \cite{Fan}.

With respect to the validity and violation of the several energy conditions associated with these regular BHs,  a detailed discussion can be found in a recent paper by Maeda    \cite{Maeda}.

\subsection{GR coupled with  Non Linear Electrodynamics}
For sake of completeness, in the following we shortly describe regular BH solutions with the inner dS core which may be obtained coupling GR with  Non Linear Electrodynamics (NLE) \cite{AA}. We follow \cite{CCZ}. However, for a recent critical discussion concerning this approach, see \cite{Bok}, where further references can be found.

The NED gravitational model is based on the following action 
\begin{equation}
S = \int d^4 x \ \sqrt{-g} (\frac{R}{2} - 2\Lambda -\mathscr{L}(I))\,,
\end{equation}
where $R$ is the Ricci scalar, $\Lambda$ is a cosmological constant, and  $I=\frac{1}{4}F^{\mu\nu} F_{\mu\nu}$ is an electromagnetic-like tensor and $\mathscr{L}(I)$ is a suitable function of it. Recall that $F_{\mu\nu}=\partial_\mu A_\nu - \partial_\nu A_\mu$. We will only deal with gauge invariant quantities, and  we put $\Lambda=0$, because its contribution can be easily  restored. The equations of motion read
\begin{equation}
\label{eq1}
G^\nu_\mu  = -F_{\alpha\nu} \partial \mathscr{L} F^{\mu\alpha} + \mathscr{L}\delta^\nu_\mu 
\end{equation}
 \begin{equation}
\label{eq2}
\nabla^\mu (F_{\mu\nu} \partial_I \mathscr{L}) = 0\,.
\end{equation}
Another equivalent approach is called dual P approach and it is based on   two  new  gauge invariant quantities 
\begin{equation}
\label{P}
P_{\mu\nu}\equiv F_{\mu\nu} (\partial_I \mathscr{L}(I))\,\,\,, P \equiv \frac{1}{4} P_{\mu\nu}P^{\mu\nu}\,,
\end{equation}
and 
\begin{equation}
\label{H}
 \mathscr{H} \equiv 2I(\partial_I \mathscr{L}(I)) - \mathscr{L}(I) \,,
\end{equation}
\begin{equation}
\label{eq_motion_2P}
\nabla^\mu P_{\mu\nu} = 0\,.
\end{equation}
In the following, we shall make use only of the traditional approach based on equations (\ref{eq1}) and (\ref{eq2}). 

Within the static spherically symmetric ansatz and from  (\ref{eq2}), one has 

\begin{equation}
\label{4}
\partial_r\left(r^2 \partial_I \mathscr{L}F^{0r}\right)   = 0\,.
\end{equation}
Since $I=\frac{1}{2}F_{0r}F^{0r}=- \frac{1}{2}F_{0r}^2$, one gets
\begin{equation}
\label{r111}
r^2 \partial_I \mathscr{L}   = \frac{Q}{\sqrt{-2 I}}\,,
\end{equation}
$Q$ being a constant of integration. As a result, within this  NED approach,  one may solve the generalized Maxwell equation. We shall make use of this equation and the the (t,t) component of the Einstein equation, which reads
\begin{equation}
\label{r}
G^t_t  =\frac{rf'+f-1}{r^2} = 8\pi\left( -2 I\partial_I \mathscr{L}+ \mathscr{L}\right)=-8\pi \rho\,.
\end{equation}
Introducing the  more convenient quantity  $X$
\begin{equation}
\label{x   }
X=Q\sqrt{-2  I}\,,          
\end{equation}
one may rewrite equation (\ref{r111}) as
\begin{equation}
\label{r2}
r^2 \partial_X \mathscr{L}   = 1\,.
\end{equation}
Furthermore, we have
\begin{equation}
\label{r4}
 \rho= X\partial_X \mathscr{L} - \mathscr{L}=\frac{X}{r^2}-\mathscr{L}  \,.
\end{equation}
Thus, when  $ \mathscr{L}(X) $ is given, then making use of (\ref{r111}), one may obtain $\rho=\rho(r)$.
 
As an example of this direct  approach, let us investigate the following class of NED models

\begin{equation}
\label{ln}
 \mathscr{L}(X)=\frac{1}{\alpha \gamma}\left((1-\frac{\gamma}{2} X^2)^{\alpha} -1\right)\,, \quad  
\partial_X \mathscr{L}=-X (1-\frac{\gamma}{2} X^2)^{\alpha-1} \,.
\end{equation}
in which $\alpha$ and $\gamma$ are two  parameters. Note that for small $\gamma$ one has the  Maxwell Lagrangian. 

Let us show that for $\gamma$  not vanishing, we may obtained  specific exact black hole solutions.
In fact, from (\ref{ln}) one gets
\begin{equation}
\label{g1}
(1-\frac{\gamma}{2} X^2)^{-2\alpha+2}=r^4 X^2\,.
\end{equation}
Again, for $\alpha=1$ one has the usual Maxwell case. Thus, we consider $\alpha \neq 1$.

The choice $\alpha=\frac{1}{2}$ leads the well known  Einstein-Born-Infeld case. With this choice one has
\begin{equation}
\label{g2}
X^2= \frac{1}{r^4+\frac{\gamma}{2} }\,.
\end{equation}
This means that the static electric field is regular at $r=0$, Born-Infeld model. Furthermore, since

\begin{equation}
\label{g22}
\mathscr{L}=- \frac{2}{\gamma} \left( 1+r^2 X \right)\,,
\end{equation}
 the effective density reads 
\begin{equation}
\label{g33}
r^2 \rho(r)=\frac{2 r^2}{\gamma}+\frac{ 2 \sqrt{r^4+\frac{\gamma}{2}}}{\gamma}   \,.
\end{equation}

In order for this object to satisfy the Weak Energy Condition) WEC, it is necessary to require $\gamma > 0$, since  $\rho$ is ill defined in the  limit $\gamma \rightarrow 0$, and  no solution associated with a vanishing electromagnetic field might exist. However, since $\gamma$ is an external parameter and not an integration constant, there is no trouble in fixing it to be positive, so that Lagrangian (\ref{ln}) satisfies the WEC.

When $\gamma >0$, the   solution is
\begin{equation}
A(r)=1-\frac{8\pi }{r}\int_0^r r^2_1\rho(r_1) dr_1 \,,
\end{equation}
and it may be expressed in term of Elliptic function, but it is easy to show there is no strictly  de Sitter core for $r \rightarrow 0$
\begin{equation}
\label{g44}
 \lim_{r \rightarrow 0}  r^2\rho(r)=\sqrt{2/\gamma}+\frac{2r^2}{\gamma}+O(r^4)  \,.
\end{equation}
The presence of the non vanishing constant $ \sqrt{2/\gamma} $ means that a conical singularity is present.

As last example, let us consider the generalized Maxwell Lagrangian

\begin{equation}
\label{chin1}
 \mathscr{L}(X)=-\frac{1}{\xi^2}\left(\sqrt{b}-\sqrt{ X} \right)^2\,, \quad  b=\frac{a}{4\pi \xi^2} \,,
\end{equation}
with $a$ and $\xi$ given parameters. One easily gets

\begin{equation}
\label{c2}
 \sqrt{X}=\frac{\sqrt{b}r^2}{r^2+\xi^2}, \quad  
\rho=\frac{b}{r^2+\xi^2} \,.
\end{equation}

Here the WEC is satisfied as long as $b$ is positive. However, also in this case $B$ is just an external parameter; its being positive rests on the condition $a \geq 0$, but this is a safe condition, since we are allowed to impose it into the Lagrangian.

For this model, one has no conical singularity in the origin, namely a de Sitter core, and the particular solution reads
\begin{equation}
\label{g4}
A(r)=1-\frac{2a}{\xi^2}+ \frac{2a}{\xi r} \arctan (\frac{r}{\xi})\,.
\end{equation}
However, this particular solution  is not for large $r$ asymptotically Minkoskian.

\subsection{The covariant Sakarov Criterion\label{Sakarov}}

Let us work in a generic SS space-time. We have already remarked that the areal radius $r$ is a scalar quantity, as well as $\chi=\gamma^{ab}\partial_ar\partial_b r$ in (\ref{chi0}). Thus, one may introduce another invariant, 
\begin{equation}
Z=\frac{1-\chi}{r^2}\,.    
\end{equation}
In Ref. \cite{CCZ}, it has been proposed a so called Sakarov (covariant) Criterion which states that a sufficient condition to deal with a generic non singular SS is to assume $Z$ and its covariant derivatives uniformly bounded  everywhere. In the static case, this, for small
values of $r$ and in the the Schwarzschild gauge with $A(r)=B(r)$, leads to $A(r)=1+cr^2+..$, namely to the existence of a dS core. We have already observed that this condition renders the curvature invariants finite at $r=0$. 

We observe that on SSS space-time where $\chi=B(r)$, $Z$ is nothing else than the g-function (\ref{gfunc}). Then, by assuming a power series expansion in $r$, we get
\begin{equation}
Z(r)=g(r)=\sum_0^\infty g_n r^n = g_0 +g_1r+g_2r^2+g_3r^3+.... \,.
\end{equation}
For the sake of simplicity, we consider here only the presence of a dS core, namely one has
$g_0 \neq 0$. Furthermore, in order to have an asymptotically flat SSS , for large $r$, we know  one has to assume
\begin{equation}
g(r\rightarrow \infty) = \frac{2M}{r^3}  \,.  
\end{equation}
One may further test the regularity of the SSS solution one is dealing with, checking the regularity of other invariants associated with $Z$. All the invariants built with the contractions of the 4-vector $\nabla_\mu Z$ must be regular at $r=0$.
Let has consider the invariant $Z_1=g^{\mu \nu}\nabla_\mu \nabla_\nu Z$, namely the d'Alambertian of $Z$.
One has (here, $A(r)=B(r)$)
\begin{equation}
Z_1=\frac{2 g'(r)}{r} -2rg(r)g'(r)+A'(r)g'(r)+A(r)g''(r)=   \frac{2 g'(r)}{r} +Z_{12}\,.
\end{equation}
Here $Z_{12}$ is a smooth function of $r$. The first term  reads
\begin{equation}
    \frac{2 g'(r)}{r} =\frac{2 g_1}{r}+4g_2+6g_3r+... 
\end{equation}
As a result,  one has to deal with $ g_1=0$, thus the scalar $Z_1$ is uniformly bounded for every $r$. 

We can continue  considering the invariant  $Z_2=g^{\mu \nu}\nabla_\mu \nabla_\nu Z_1$, and we get
\begin{equation}
Z_2= \frac{2 Z_1'(r)}{r} +Z_{22}\,.
\end{equation}
Here $Z_{22}$ is a smooth function of $r$, while the first term  reads
\begin{equation}
    \frac{2 Z_1'(r)}{r} = \frac{12g_3}{r}+O(r)\,. 
\end{equation}
As a result, if we require the scalar $Z_2$ to be uniformly bounded for every $r$, one should assume 
\begin{equation}
g_3=0\,.    
\end{equation}
Continuing in this way,  requiring $Z_n=(g^{\mu \nu}\nabla_\mu \nabla_\nu)^n Z$ to be uniformly bounded for every $r$, it follows that $g_{2n+1}=0$, namely $g(-r)=g(r)$.  This result is in agreement with the one obtained recently in Ref. \cite{Giac}. Thus, a BH admitting a dS core is regular only if the g-function is an even function of $r$. 

Let us consider some explicit examples. The well known 
Hayward BH (\ref{h2}) does not fulfill this requirement, as it is also stressed in Ref. \cite{Giac}, where other BHs examples have been investigated.  

In our example of delta-like regularized BHs (\ref{gG}),  one has to take $a$ even number (see (\ref{gGsmallr})). 
Thus, the relate g-functions are even functions in $r$. For example, the Bardeen and Dynmikova BHs belong to this class of regular BHs.

Finally, we mention the class of regular BHs, the one with an inner Minkowsky core, recently re-proposed by Simpson and Visser \cite{Culetu,Culetu2, SVM, SVM2}. In the Schwarzschild gauge, the metric reads,
\begin{equation}
ds^2=-\left(1-\frac{2m e^{-\ell/r}}{r}\right)dt^2 +\frac{dr^2}{\left(1-\frac{2m e^{-\ell/r}}{r}\right)}+r^2dS^2\,,  
\end{equation}
$\ell\,,m$ constants.
The g-function associated with this class reads
\begin{equation}
g(r)=\frac{2M}{r^3}e^{-\frac{\ell}{r}}   \,.
\end{equation}
When $r \rightarrow 0$, all the curvature invariants vanish.

One might think that the further requirement we have stressed to adopt in order to deal with a regular BH having an inner dS core, namely $g(-r)=g(r)$, depends on the fact we are working in the Schwarzschild gauge.  In fact, for example, performing the coordinates change $r=\sigma^2$, one obtains
\begin{equation}
  ds^2=-A(\sigma^2)dt^2+\frac{4\sigma^2 d\sigma^2}{A(\sigma^2)}+\sigma^4 dS^2\,.  
\end{equation}
In this way all the components of the metric are always even function of $ \sigma$, and also the invariant $ Z=g(\sigma^2)$. However, if $g(r)$ is not an even function, the divergences in the invariant are still present.

At this point, we may take an example, namely the BH in Ref. \cite{Fan} with
\begin{equation}
A(r)=B(r)=1-\frac{r^2}{(r+\varepsilon)^3}\,,    
\end{equation}
such that 
\begin{equation}
g(r) =\frac{1}{(r+\varepsilon)^3}\,, 
\end{equation}
is not an even function and we expect the divergence in $Z_1$.

After the identification $r=\sigma^2$ we have
\begin{equation}
g(\sigma^2) =\frac{1}{(\sigma^2+\varepsilon)^3} \,.
\end{equation}
Now, if we compute the invariant   $Z_1=g^{\mu \nu}\nabla_\mu \nabla_\nu Z$, we derive
\begin{equation}
Z_1=\frac{1}{4\sigma^2}\left( 3\frac{g'(\sigma^2)}{\sigma}+g''(\sigma^2) \right)+O(\sigma)\,.    
\end{equation}
Therefore,
\begin{equation}
 Z_1=-\frac{6}{\sigma^2} \frac{1}{\varepsilon^4}+O(\sigma)\,,
\end{equation}
and we see that the result in the Schwarschild gauge is still valid: if $g(r)$ is not even function, there exists a divergent invariant when $r \rightarrow 0$.

\section{ Alternative Regular Black holes: Black Bounce space-times}

 These regular space-times have been dubbed ``Black Bounce space-times'' by Visser et al. \cite{SV, SV2, SV3}. In the examples of static metrics we are going to discuss,   no interior dS core is present and the central singularity is avoided thanks to the introduction of a minimal length scale.
 
 The starting point is the following metric
\begin{equation}
ds^2=-A(r)dt^2+\frac{dr^2}{A(r)}\frac{1}{(1-f_\ell(r))}+r^2 d\Omega^2   \,.
\end{equation}
with $f_\ell(r)$ positive function of $r$ such that $f_\ell(r) \rightarrow 0$ for 
$\ell \rightarrow 0$ and $f_\ell(r) \rightarrow 0$ for $r \rightarrow\infty$. Therefore,
 $\ell$ is a small length parameter, for example on the order of the Planck length.

Since in the Schwarzschild gauge (\ref{sg00}) the metric function $B(r)$ is given by, 
\begin{equation}
B(r)=A(r)(1-f_\ell(r)) \,,
\end{equation}
it follows  that the range of $r$ is  restricted by the condition 
$\frac{A(r)}{B(r)}>0$, namely  $0<f_\ell(r)<1$.
The possible horizons  satisfy
\begin{equation}
A(r)(1-f_\ell(r))=0  \,,
\end{equation}
namely are located at $r=r_H, r_0$ such that
\begin{equation}
A(r_H)=0\,,\quad f_\ell(r_0)=1\,.    \end{equation}
When $r_0$ is smaller than $r_H$, one has a BH and $r_0$ is identified with a sort of minimal length scale for the metric. In the other case, one is dealing with  WH, $r_0$ represents the WH throat and $r_H$ may be also absent. Note that in general $r_0\equiv r_0(\ell)$.

Within $F(R)$-modified gravity models with the on-shell condition $F(0)=F_R(0)=0$,  an attempt to deal with such BHs has been presented by Bertipagani et al. \cite{Berti}, but in this case  a Cauchy horizon is generically present unless one makes a specific choice for $\ell$.

We observe that in the  the Eddington-Filkenstein (EF) gauge, one has
\begin{equation}
ds^2=-A(r)dv^2+2\sqrt{\frac{1}{(1-f_\ell(r))}}dv dr+r^2d\Omega^2   \,,
\end{equation}
and the singularity at $(1-f_\ell(r))$ is harmless and may be removed by a suitable change of coordinates.

Coming back to the framework of GR in the presence of fluid, the simplest and physically relevant choice for $A(r)$ is given by,
\begin{equation}
A(r)=1-\frac{C}{r}\,, \quad C=2m\,,
\end{equation}
where $m$, as usually, is a constant mass parameter.
Thus, making use of Eq. (\ref{EOM1}),
 the associated energy density reads
\begin{equation}
\rho=\frac{ f_\ell(r)+(r-C)f_\ell'(r) }{8\pi r^2}\,.
\end{equation}
From Eq. (\ref{EOM2}) one derives,
\begin{equation}
p_r=- \frac{ f_\ell(r) }{8\pi r^2} \,,
\end{equation}
while from Eq. (\ref{TOeq}) we get
\begin{equation}
p_T=\frac{(C/2-r)f_\ell'(r) }{16\pi r^2}  \,.
\end{equation}
Thus,
\begin{equation}
\rho+p_r=\frac{(r-C)f_\ell'(r) }{8\pi r^2}\,.
\end{equation}
As a consequence, the Null Energy Condition (NEC) may be violated, but due to the restrictions $1>f_\ell(r)>0$, $r>r_0(\ell)$, all the physical quantities and curvature invariants are  bounded. For example the Ricci curvature results to be
\begin{equation}
R=16 \pi \rho+\frac{C f_\ell'(r) }{2 r^2}\,.
\end{equation}  
Let us see some examples of these non-singular BH metrics.
The simplest choice for $f_\ell(r)$ is 
\begin{equation}
f_\ell(r)=\frac{\ell}{r}\,.\label{f00}
\end{equation}
This has been used by D'Ambrosio-Rovelli \cite{DR, DR2} and Bertipagani et al \cite{Berti}. The  static version of  D'ambrosio- Rovelli solution  reads
\begin{equation}
ds^2=-\left(1-\frac{C}{r}\right)dt^2+\frac{ dr^2}{(1-C/r)(1-\ell/r)}+r^2d\Omega^2\,.\label{AR}
\end{equation}
The Simpson-Visser choice is \cite{SV}
\begin{equation}
f_\ell(r)=\frac{\ell^2}{r^2}\,,
\end{equation}
with the related metric,
\begin{equation}
ds^2=-\left(1-\frac{C}{r}\right)dt^2+\frac{ dr^2}{(1-C/r)(1-\ell^2/r^2)}+r^2dS^2\,.
\label{SVBH}
\end{equation}
In both cases, we have the restriction on $r$ as $r>\ell$.
The possible horizons are $r_H=C$ and
 $r=\ell$. When $\ell <C$, one has a BH, for
$\ell>C$ one is dealing with a WH. These metrics are  regular for $\ell>0$, namely  $r> \ell$, and the singularity in $r=0$ is  avoided.

In order to remove the coordinate singularity, it is   convenient to make use of another radial coordinate, namely $r=\sqrt{\sigma^2+\ell^2}$, with $-\infty <\sigma< +\infty$. 
With regard to this new radial coordinate, the most natural choice for $f_\ell$ is \cite{SV}
\begin{equation}
f_\ell(r)=\frac{\ell^2}{r^2}\,,
\end{equation}
and we get
\begin{equation}
ds^2=-A(\sigma)dt^2+\frac{d\sigma^2}{A(\sigma)}+
(\sigma^2+\ell^2)d\Omega^2\,.
\end{equation}
With the choice (\ref{f00}) the metric  reads, in terms of $\sigma$
\begin{equation}
ds^2=-A(\sigma)dt^2+\frac{d\sigma^2}{A(\sigma)}\left(1+\frac{\ell}{\sqrt{\sigma^2+\ell^2}}\right)+
(\sigma^2+\ell^2)d\Omega^2\,.
\end{equation}
For a generic metric, one as a regular BH as soon as the function  $A(\sigma)$ and its derivatives with respect to $\sigma$ are everywhere finite.

For example, the Simpson-Visser BH (\ref{SVBH}) assumes the simple form,
\begin{equation}
ds^2=-\left(1-\frac{C}{ \sqrt{\sigma^2+\ell^2}}\right)dt^2+\frac{d\sigma^2}{\left(1-\frac{C}{\sqrt{\sigma^2+\ell^2}}\right)}+
(\sigma^2+\ell^2)d\Omega^2\,.
\end{equation}
The horizon is given by $\sigma_H=\sqrt{C^2-\ell^2}$, and for large $\sigma$, one gets $A(\sigma)=B(\sigma)=1-\frac{C}{\sigma}-\frac{C\ell^2}{2\sigma^3}\times...$

For the D'ambrosio- Rovelli metric (\ref{AR}) one has,
\begin{equation}
ds^2=-\left(1-\frac{C}{ \sqrt{\sigma^2+\ell^2}}\right)dt^2+\frac{d\sigma^2}{\left(1-\frac{C}{\sqrt{\sigma^2+\ell^2}}\right)}\left(1+\frac{\ell}{\sqrt{\sigma^2+\ell^2}}\right) +(\sigma^2+\ell^2)d\Omega^2\,,
\end{equation}
the horizon is given by $\sigma_H=\sqrt{C^2-\ell^2}$, and for large $\sigma$ we have $A(\sigma)=1-\frac{C}{\sigma}-\frac{C\ell^2}{2\sigma^3}+..$, and $B(\sigma)=1-\frac{C+\ell}{\sigma}-\frac{C\ell^2+\ell^3}{2\sigma^3}+..$.

A further example is the Peltola-Kunstetter BH,  motivated by Loop Quantum Gravity (LQG) \cite{PK}. The related space-time is given by,
\begin{equation}
ds^2=-\left(\sqrt{1-\ell^2/r^2}-\frac{C}{r}\right)dt^2+\frac{ dr^2}{\left(\sqrt{1-\frac{\ell^2}{r^2}} -C/r\right)(1-\ell^2/r^2)}+r^2dS^2\,,
\end{equation}
 or, in the free coordinate singularity form,
\begin{equation}
ds^2=-\frac{(\sigma-C)}{ \sqrt{\sigma^2+\ell^2}}dt^2+\frac{d\sigma^2}{\frac{(\sigma-C)}{\sqrt{\sigma^2+\ell^2}}}+
(\sigma^2+\ell^2)d\Omega^2\,.
\end{equation}
 The horizon is located at $\sigma_H=C$, and and for large $\sigma$ one has $A(\sigma)=B(\sigma)=1-\frac{C}{\sigma}--\frac{\ell^2}{\sigma^2}-\frac{C\ell^2}{2\sigma^3}+...$

Related to these examples there is the BH solution found by Modesto within Loop Quantum gravity (LQG) approach (see ref. \cite{Mo} and references therein ).

 These metrics are quite interesting examples of regular BHs, asymptotically flat,  without Cauchy horizon.  
 The issue of their stability can be investigated by studying the Quasi Normal Modes (QNMs).
 The other important issue is the generalization to the rotating case.
 For the Simpson-Visser BH, this has been done and the result is the Kerr rotating metric with the new radial  $\sqrt{\sigma^2+\ell^2}$ replacing $r$ \cite{Mazza, Mazza2}.  With regard to this issue, the possible physical relevance for these metrics has been subjected to several investigations, see  \cite{Si}, where further references can be found.
 
 It should be noted that some of these metrics can be derived by suitable Non Polynomial Lagrangian (NPL) (see Ref. \cite{CCZ} where it is also discuss a class of metrics describing regular BHs without the Cauchy horizon issue).

\section{Black holes in four dimensional regularized Lovelock models}

Recently there has been a increasing interest in an approach initiated by Tomozawa \cite{To},  further revisited and extended to flat Friedman-Lema\^itre-Robertson-Walker (FLRW) cosmological models in \cite{CSZ}, and recently rediscovered and extended in \cite{GL}. 

Several aspects have been considered in \cite{A, A2}. It should be stressed that such procedure has been criticized in literature, see for example \cite{Arr}. However, in Ref. \cite{Clifton22} one can found   a recent and very complete review, containing a vast bibliography on this issue. 

The basic idea is to bypass the Lovelock theorem which states that the only gravitational theory admitting second order equations of motion in four dimension ($D=4$) is  the Einstein-Hilbert, plus a cosmological constant term. 
On the other hand, in $D>4$, the so called Lovelock contributions are possible, and the equations of motion are still of second order. However, by making use of a suitable regularization procedure, it is possible to include non-trivial Lovelock additional contributions in $D=4$ also.

Let us start with the simplest case.
The starting point is the following gravitational action in the generic dimension $D$,  
\begin{equation}
I =\int d^Dx \sqrt{-g}   \left(R -\xi \frac{G}{D-4}\right)\,,
\end{equation}
where  $\xi$ is real  and $G$ is the Gauss-Bonnet Lovelock contribution, namely
\begin{equation}
G=R_{\mu \nu \alpha \beta}R^{\mu \nu \alpha \beta}-4R_{\mu \nu }R^{\mu \nu} +R^2\,.
\end{equation}
It is well known that in $D=4$ the Gauss-Bonnet is a topological invariant and it does not contribute to the equations of motion. 

The trick (motivated by dimensional regularization), consists in having included in the action the factor $\frac{1}{D-4}$, regularizing the Gauss-Bonnet coupling constant. As a result, evaluating the equations of motion in a $D$-dimensional SSS space-time, and then taking the limit $D \rightarrow 4$, one has for 
four dimensional SSS space-time (\ref{sg00})
the solution \cite{To,CSZ,GL},
\begin{equation}
A(r)=B(r)=1-\frac{r^2}{\xi}\left(1-\sqrt{1-\frac{8\xi m}{r^3}}   \right)  \,,  
\label{GBsol}
\end{equation}
where $m$ is an integration constant and
we assume $\xi >0$. The solution represents a BH because $A(r)=0$ gives only a positive root, 
\begin{equation}
r_H=m+\sqrt{m^2+\xi}\,.  
\end{equation}
We note that the other root
$r=m-\sqrt{m^2+\xi}$
is negative when  $ \xi>0$ and there is no Cauchy horizon. Furthermore, for large values of $r$, one has $A(r)=1-\frac{2m}{r}+...$ and $m$ can be identified with the mass of BH. 

Finally, there exists a restriction on the values of $r$, since the solution above is real as soon as \cite{To},
\begin{equation}
r>r_c= 2 (m\xi)^{1/3}\,.    
\end{equation}
Therefore, one gets an  asymptotic flat BH without the Cauchy horizon problem. One may take $\xi=\gamma m^2$,  with the dimensionless  parameter $\gamma <<1 $. The critical radius results to be $r_c=2m(\gamma)^{1/3}$,
and the horizon is located at
\begin{equation}
r_H=m(1+\sqrt{1+\gamma})\,.  
\end{equation}
However, also with the above restriction, since the derivatives of $A(r)$ are ill defined at the critical radius, if follows that the curvature invariants are ill defined  there, and one is not dealing with a regular BH.

This regularization procedure may be generalized to higher order Lovelock gravity, including a suitable dimensional dependent factor in all higher order Lovelock coupling constants \cite{kun,Aim,cine}.
Again, the starting point is the following gravitational action in $D$-dimension,  
\begin{equation}
I =\int d^Dx \sqrt{-g}   \left(R +\sum_p  a_pL_p\right)\,,
\end{equation}
in which there is no cosmological constant and the Einstein-Hilbert action leads to $a_p=0$. All the other higher curvature terms may be dimensionally regularized   including a suitable factor  $ (\prod_(k D-k))^{-1}$ in all higher order Lovelock coupling constants \cite{kun}, and   $L_p$ are the related Lovelock higher curvature invariants. 

In the limit $D\rightarrow 4$, and in the vacuum, the solution of the regularized EOMs  is
\begin{equation}
 \frac{C}{r^3} = \sum_p c_p g(r)^p=G(r)\,,\quad g(r)=\frac{1-A(r)}{r^2}\,,
\end{equation}
with $g(r)$ function of $r$ and
$C$ the integration constant of the solution.
 When the sum is over a  finite number depending on $D$,  one  may determine $g(r)$, solving an algebraic equation, and then $A(r)$ is found. In general the BH solutions one gets are singular.  For example, in GR,  $G(r)=g(r)=\frac{C}{r^3}$, and one has   $A(r)=1-\frac{C}{r}$. In the Gauss-Bonnet case,  $G(g)=g(r)-\xi g(r)^2$ and one finds the regularized Gauss-Bonnet model discussed above, since 
\begin{equation}
\frac{C}{r^3}=g(r)-\xi g(r)^2 \,,   
\end{equation}
 solution being
\begin{equation}
g(r)=\frac{1}{2\xi}\left(1-\sqrt{1-\frac{4\xi C}{r^3}}\right)\,,
\end{equation}
and we recover (\ref{GBsol}) once $\xi\rightarrow \xi/2$.
However, we may consider an  infinite number of suitable dimensional regularized Lovelock terms \cite{kun,Aim,cine}, and if the arbitrary coupling constants left are properly chosen, the  sum may be considered the expansion of a function $G(r)\equiv G(g(r))$ within its radius of convergence, and one derives
\begin{equation}
 \frac{C}{r^3} =G(g(r))\,\quad g(r)=\frac{1-A(r)}{r^2}\,.
\end{equation}
As an interesting application within this infinite regularized Lovelock models,  we  present  two non trivial examples. The first one is related to the choice
\begin{equation}
G(g(r))=\frac{2M h(r)}{(1-\varepsilon^2 h(r)^{2/3})^{3/2}}  \,, \quad h(r)=\frac{g(r)}{2M}\,,
\end{equation}
$\varepsilon$ being a constant parameter arbitrarily small.
Thus,
\begin{equation}
 h(r)=\frac{1}{(r^2+\varepsilon^2)^{3/2}} \,,   
\end{equation}
and one gets the Bardeen BH,
 \begin{equation}
A(r)=1-\frac{2M r^2}{( r^2+\varepsilon^2)^{3/2}  }  \,.   
 \end{equation}
The second choice is \cite{kun,cine}
\begin{equation}
G(g(r))=\frac{h(r)}{1-b^3 h(r)}  \,,
\quad h(r)=\frac{g(r)}{2M}\,,
\end{equation}
$b$ being a constant.
Thus,
\begin{equation}
 h(r)=\frac{1}{r^3+b^3} \,,   
\end{equation}
and one gets  
the Poisson-Israel BH (\ref{h1}) with $b^3=\ell^3$ or 
Hayward BH (\ref{h2}) with $b^3=\ell^2 m$.
Thus, we have shown that Bardeen and Hayward and other regular BHs may be also derived as vacuum solutions within this Lagrangian framework.

\section{Non-singular cosmological models}

In this section we would like to discuss some non singular cosmological models. 
We will work within flat FLRW models whose metric reads
\begin{equation}
ds^2=-dt^2+a(t)^2 d\Vec{x}^2\,,    
\end{equation}
where $a\equiv a(t)$ stands for the expansion factor and is a function of the cosmological time $t$. Moreover, in what follows we will use the Hubble parameter $H\equiv H(t)=\frac{\dot a(t)}{a(t)}$, where the dot is the derivative with respect to the time.
In GR the first Friedmann equation and matter conservation law read,
\begin{equation}
3H^2=8\pi \rho\,,
\end{equation}
\begin{equation}
 \frac{d \rho}{dt}+3H(\omega+1)\rho=0\,,
\label{conslaw}
\end{equation}
where $\rho\,,p$ are the energy density and pressure of matter contents of the Universe, respectively, and we are assuming the EoS
$p=\omega \rho $ with the EoS parameter $\omega$ constant. As well known,  the matter conservation has as solution  $\rho=\rho_0 a^{-3(1+\omega)}$ ($\rho_0$ is the energy density when $a(t)=1$) and,
if $\omega\neq -1$,
these equations lead to the well known Big-Bang (or Big-Rip) singularity when $a(t)$ vanishes.

If we apply the Covariant Sakarov Criterion (see \S \ref{Sakarov}), we have 
\begin{equation}
\chi=1-H^2\,, \quad Z=\frac{1-\chi}{r^2}=H^2\,.
\end{equation}
Thus, one has to deal with $Z$ and therefore $H(t)$ and its derivatives and to require they are uniformly bounded. In a FLRW Universe, if $H$ and $\dot{H}$ are  uniformly bounded, then the space-time is  causally geodesically complete (see Ref.\cite{ifi} and reference therein).

However, in  this cosmological framework, we may investigate the absence of singularities working directly on the equations of motion as follows. 

Loop Quantum Cosmology (LQC) \cite{Bo} (see also \cite{Barca}), modified gravity and mimetic gravitational models\cite{CA20}, NPL  models \cite{CCZ1}, Lovelock regularized models \cite{kun,Aim,cine} lead to a first generalized Friedmann equation of the type
\begin{equation}
3H^2=F(\rho)    \,,
\label{1F}
\end{equation}
or
\begin{equation}
G(H^2)=\frac{\rho}{3}    \,,
\label{2G}
\end{equation}
where the form of $F(\rho)$ and $G(H^2)$ 
depends on the model under consideration and on shell are functions of $\rho$ or $H^2$. Moreover,
the matter conservation law (\ref{conslaw}) is still valid.

\subsection{The  $3H^2=F(\rho)  $ case}

When  $F(\rho)$  is a positive or negative function, in general the Big-Bang singularity is  present. This may be understood as a consequence of the Osgood Criterion (OC) (see for example Ref. \cite{Koh}): 
given $y\equiv y(t)$,
if $f(y)$ is never vanishing (always positive or negative) and one has the initial value problem
\begin{equation}
\dot{y}=f(y(t))\, \quad y(t_0)=y_0\,,    
\end{equation}
then there exists a finite singularity iff
\begin{equation}
\int_{y_0}^\infty \frac{dy}{f(y)}<\infty\,.
\end{equation}
In our case from Eq. (\ref{1F}) and Eq. (\ref{conslaw}) we get
\begin{equation}
\dot{\rho}=-\sqrt{3 F(\rho)}(1+\omega)\rho \,,
\end{equation}
and if $F(\rho)>0$ we do not have any restriction on the range of $\rho$.
We assume $\omega\neq -1$ and $\rho(t_0)=\rho_0$. 
Now if
\begin{equation}
\frac{1}{\sqrt{3}(1+\omega)}
\int_{\rho_0}^\infty \frac{d\rho}{ \rho \sqrt{F(\rho)}}<\infty\,,
\end{equation}
the OC guarantees the existence of a finite-time singularity.
For example, the string brane correction leads to
\begin{equation}
3H^2=\rho+\alpha \rho^2    \,,\quad \alpha>0\,.
\end{equation}
Thus, $F(\rho)= (\rho+\alpha \rho^2)$, no restriction on $\rho$ is a priori present,  $    \frac{1}{ \rho \sqrt{F(\rho)}} $ is summable, and OC leads to the existence of a finite-time singularity. One derives the exact solution
\begin{equation}
\rho=\frac{\rho_0}{-\alpha+  \frac{3}{4}(1+\omega)^2t^2 } \,,   
\end{equation}
and the related scale factor results to be,
\begin{equation}
a(t)=\left(-\alpha+\frac{9}{4}(1+\omega)^2t^2\right)^{\frac{1}{3(1+\omega)}}  \,.  
\end{equation}
As a consequence,  $a(t)$ is vanishing at $t^2_s=\frac{4\alpha}{3(1+\omega)^2}$ , and $\rho$ and $H^2$ diverge there. For $\alpha=0$ we recover the GR case. If $\omega>-1$ the singularity is present at $t=0$, otherwise $a(t)$ is well defined by replacing $(1+\omega)t\rightarrow -(1+\omega)(t^*-t)$, where $t^*$
is an integration constant corresponding to the time of future singularity \cite{Caldwell}.

The thing changes in the case of Loop Quantum Cosmology (LQC) or other similar cases. In fact, one has
\begin{equation}
3H^2=\rho-\alpha\rho^2    \,,\quad \alpha>0\,.
\end{equation}
Now,  since $H^2>0$, the OC cannot be  applied, due to the the constraint $\rho <\frac{1}{\alpha}=\rho_c$, and the exact solution confirms the non-existence of the time singularity and reads,
\begin{equation}
a(t)=(\alpha+\frac{3}{4}(1+\omega)^2t^2)^{\frac{1}{3(1+\omega)}}   \,. 
\end{equation}
Here, $a(t)$ is always positive, and $\rho$ and $H^2$ are bounded. Thus, all the curvature invariants are bounded.

In general, when $F(\rho)$ is not positive definite and admits a (positive) fixed point, one may argue as follows \cite{CA20}. Let us rewrite equation (\ref{1F}) as
\begin{equation}
\left(\frac{d a}{dt}\right)^2=\frac{a^2}{3}F(\rho(a))=Y(a)\,.    
\end{equation}
We assume that there exists a positive fixed point $F(\rho(a_*))=0\,,a_*=a(t_*) $, namely $Y(a_*)=0$. By expanding $Y(a)$ near to this fixed point we obtain,
\begin{equation}
Y(a)=\frac{d Y(a^*)}{d a}(a-a_*)\,.
\end{equation}
Solving the differential equation above, one gets
\begin{equation}
a(t)=a_*+\frac{1}{4}\frac{d Y(a^*)}{d a}t^2  \,.  
\end{equation}
The chain rule together with the solution of conservation law (\ref{conslaw}), $\rho=\rho_0 a^{-3(1+\omega)} $, lead to
\begin{equation}
a(t)=a_*-\frac{(1+\omega)}{4}\,  a_*^{-(2+3\omega)}\, \left(\frac{\partial F}{\partial \rho}\right)_{t_*}\, t^2\,.
\end{equation}
As a result, if $ \left(\frac{\partial F}{\partial \rho}\right)_{t_*}<0 $, there is no Big Bang-like singularity.

As a check,  in the case of LQC, $F(\rho)=\rho-\alpha \rho^2 $ with a fixed point $\rho(a_*)=\frac{1}{\alpha}$, and $ \left(\frac{\partial F}{\partial \rho}\right)_{t_*}=-1  $, such that there is no Big Bang singularity.

A simple generalization is given by $F(\rho)=\rho-\alpha \rho^2-\beta \rho^3 $, with $\alpha>0 $ and $\beta >0$. The positive fixed point is 
\begin{equation}
\rho(a_*)=\frac{-\alpha+\sqrt{\alpha^2+4\beta}}{2\beta}\,,
\end{equation}
and
\begin{equation}
 \left(\frac{\partial F}{\partial \rho}\right)_{t_*}=-\rho_*\left(\alpha+2\beta \rho_* \right) <0\,.  
\end{equation}
Again, no Big Bang singularity is present.

\subsection{The  $G(H^2)=\frac{\rho}{3}$ case }

As in the previous case, when $G(H^2)$ is always positive, the Big Bang like singularity is present.
In fact, we may again argue as follows.
First, taking the derivative with respect to $t$ of equation (\ref{2G}) and making use of the matter conservation law (\ref{conslaw}), one obtains
\begin{equation}
\frac{d H}{dt}=-\frac{3(1+\omega)}{2}\frac{G(H^2)}{G_{H^2}(H^2)}=Y(H^2)\,, \quad G_{H^2}(H^2)=\frac{\partial G(H^2)}{\partial H^2}\,.   
\end{equation}
Now, by introducing $H_0^2\equiv H^2(t_0)$ and if $ \frac{G(H^2)}{G_{H^2}(H^2)}$ is positive or negative definite,  
\begin{equation}
-\frac{2}{3(1+\omega)}
\int_{H_0^2}^\infty dH^2 \frac{G_{H^2}(H^2)}{G(H^2)} <\infty\,,  
\end{equation}
then OC says that a finite-time singularity is present. For example, if $G(H^2)=H^2+\xi^2H^4$, $\xi^2$ a positive constant, the OC hypotesis are satisfied and there is the singularity.

If $G(H^2)$ is not positive definite, then  there exists a fixed point $H_*^2$ for which $G(H_*^2)=0$ and by expanding around the positive fixed point we get
\begin{equation}
\frac{d G(H^2)}{d H^2}(H^2-H_*^2)=\frac{\rho}{3}\,.  
\end{equation}
By using the fact that
\begin{equation}
\rho=-\frac{\dot\rho}{3H(1+\omega)}=-\frac{2}{3(1+\omega)}\frac{d G(H^2)}{H^2} \dot H\,,   
\end{equation}
we obtain
\begin{equation}
\dot H=-\frac{9(1+\omega)}{2} \left(H^2-H^2_*\right)\,.
\end{equation}
Thus, one has for small $t$,
\begin{equation}
H(t)=H_*\tanh\left( \frac{9(1+\omega)}{2}H_* t\right)\,.   
\end{equation}
From this it follows $H^2(t)< H^2_*$, and
\begin{equation}
a(t)=\left(\cosh\left(  \frac{9(1+\omega)}{2}H_* t)\right) \right)^{\frac{2}{9(1+\omega)}}\,, 
\end{equation}
namely the Big Bang singularity is absent.  

As an interesting example, let us consider $G(H^2)=H^2-\xi H^4$,  $\xi >0$. The fixed point is $H^2_*=\frac{1}{\xi}>0$. Thus, there is no Big Bang singularity in agreement with Ref. \cite{CSZ}.

This is a quite general result concerning approximate solution of generalized Friedmann equation, since it is a consequence only of the existence of a positive fixed point. In this case $G(H^2)$ cannot be always positive or negative. 

When $G(H^2)$ is always positive/negative, a negative/positive fixed point $H^2_*$ may exist, formally an imaginary $H_*$. One may proceed as above, making the expansion around this fixed point, and, for small $t$ ,   the solution for $a(t)$ is now

\begin{equation}
a(t)=\left(\cos\left(  \frac{9(1+\omega)}{2}|H_*| t) \right)\right)^{\frac{2}{9(1+\omega)}}   \,.
\end{equation}
As a result, a singular cyclic universe is present, in agreement with the numerical results presented in Ref. \cite{Clifton22}.

We conclude with this remark. As in the static case, we may generalize the Lovelock regularized model in $D=4$ in the cosmological setting considering an infinite number of Lovelock contributions. Thus the first generalized Friedmann equation reads
\begin{equation}
G(H^2)=\sum_pc_p H^{2p}  =\frac{\rho}{3}\,,
\end{equation}
 with $G(H^2)$ a suitable  function of $H^2$ depending on the coefficients $c_p$ (for example, $c_0=0\,,c_1=1$). 

An interesting example is given by,
\begin{equation}
G(H^2)=\frac{1}{2\alpha}    \left(1-\sqrt{1-12\alpha H^2}  \right)\,,
\end{equation}
with $\alpha>0$, which leads to LQC modified equation 
\begin{equation}
3H^2=\rho-\alpha\rho^2    \,.
\end{equation}
Other examples can be found in Ref. \cite{cine}.

\section{Conclusions}

In this paper, the problem of singularity in spherically symmetric space-times has been investigated. In the specific, some aspects of regular BHs  and non singular cosmological models have been discussed. 

We have constructed a class of regular BH solutions in the framework of GR and in the presence of a delta-like regularized source.  The metric components  have been expressed as $A(r)=B(r)=1-r^2 g(r)$, namely via the ``g-function'' $g(r)$. This kind of metrics are asymptotically flat and show a Schwarzschild-like behaviour  at large distances. 
In the case of  regular BHs with dS core the central singularity is absent. However, even though all the curvature invariants are bounded, for the class of models which do not admit even g-function, namely $g(-r) \neq g(r)$, there exist invariants built with covariant derivatives which are not bounded when $r \rightarrow 0 $. This result can be achieved  
by making use of the Sakarov Criterium which offers a  method which permits to discriminate between singularity free solutions and solutions which are not. 

We may add this remark. Regular BHs with $g(r)$ being an even function (for example Bardeen BH) admit an extension to negative values of $r$. As a consequence, the curvature invariants have singularities only for imaginary values of $r$. On the other hand, regular BH with $g(r)$ being a non even function  (for example the Hayward BH) have curvature singularities for real (negative) values of $r$. Similar ideas have been  put forward in a very recent paper by Modesto et al. \cite{Mo2}, where geodesic incompleteness of a class of regular black holes with $g(r) \neq g(-r)$ is discussed. Furthermore, this issue may be also related to the determination of the QNM asymptotics of regular BH via the monodromy approach (see Ref. \cite{Lan}). 

We should note that this first kind of regular BHs may suffer of an instability issue associated with the presence of the Cauchy inner horizon. In this respect, it has been noticed that, if the Cauchy horizon surface gravity is vanishing, then the Cauchy horizon instability may be avoided \cite{Berti,Car}. In the model investigated in Ref. \cite{Berti}, this is equivalent to the introduction of a minimal length scale in the metric, for example on the Planck scale, and the resulting metric falls in the other class of regular BHs, the so called Black Bounce space-times.

We have also showed that many regular BH solutions may be also viewed as vacuum solutions of four dimensional regularized Lovelock models. 
As a consequence, there exist  alternative Lagrangian methods as the  NPL approach \cite{CCZ} and the Lovelock regularized approach discussed in this paper, which permit to derive such regular BHs besides the Non Linear Electrodynamics approach \cite{A}.

In the final part of our work, we have presented a general approach to the problem of finite-time singularities in flat FLRW space-times cosmological models, making use of the so called Osgood Criterion.

\section{Acknowledgments}
We would like to thank Aimeric Colleaux, Massimiliano Rinaldi and Luciano Vanzo for many useful discussions.

\section*{Appendix:  A note on Painlev\`e gauge and Hawking temperature }

When one deals with SSS space-time, it may be convenient to adopt different metric gauges with respect to the Schwarzschild one. For example, 
by using the Eddington-Filkenstein gauge with 
$ \frac{A(r)}{B(r)}>0$, we can rewrite the static metric (\ref{sg00}) as        
\begin{equation}
ds^2=-A(r)dv^2+2\sqrt{\frac{A(r)}{B(r)}}dvdr+ r^2 dS^2\,.
\label{pg}
\end{equation}
Here $v$ is the advanced time coordinate and $v=cost$ represents radial in-going null geodesics.
In this coordinate system the metric is explicitly non-singular at the BH radius.

An other useful reference system where
there is no coordinate singularity at the BH event horizon is 
the Painlev\`e gauge with 
\begin{equation}
    dt=dT-\sqrt{\frac{(1-B(r))}{A(r)B(r)}} dr\,,
\end{equation}
such that the metric (\ref{sg00}) can be rewritten as
\begin{equation}
ds^2=-A(r)dT^2+2\sqrt{\frac{A(r)}{B(r)}(1-B(r))}dTdr+dr^2+ r^2 d S^2\,,
\end{equation}
and a further restriction is present since $B(r)<1 $. It means that not for every SSS metric the Painlev\`e gauge can be extended to all range of the radial coordinate $r$.

For example, static Kottler solution with \begin{equation}
A(r)=B(r)=1-\frac{2m}{r}-H_0r^2\,,
\end{equation}
$m\,,H_0$ being constants, leads to
\begin{equation}
1-B(r)=\frac{2m}{r}+H^2_0r^2>0\,,
\end{equation}
and the metric exists for any value of $r$.
However, if one considers the Schwarzshild Anti-De Sitter (AdS) BH solution with
\begin{equation}
B(r)=   1 -\frac{2m}{r}+H_0^2 r^2\,,
\end{equation}
we get
\begin{equation}
1-B(r)=\frac{2m}{r}-H^2_0r^2\,,
\end{equation}
and there is a restriction on the range of $r$.  The same fact is also present in the Reissner-Nordstr{\"o}m BH solution, where
\begin{equation}
A(r)=B(r)=1-\frac{2m}{r}+\frac{Q^2}{r^2}\,,
\end{equation}
$Q$ being a constant.
Thus,
\begin{equation}
1-B(r)=\frac{2m}{r}-\frac{Q^2}{r^2}\,,
\end{equation}
which is not positive defined for any value of $r$.

We recall a possible way to deal with this issue \cite{Mpois}. One may introduce a generalized Painlev\`e time in (\ref{sg00}) as
\begin{equation}
dT=dt+\sqrt{\frac{(1-B(r)g(r))}{A(r)B(r)}} dr\,,
\label{tpgen}
\end{equation} 
with $g(r)>0$ arbitrary function such that $(1-B(r)g(r))>0$. The  associated metric is
 \begin{equation}
ds^2=-A(r)dT^2+2\sqrt{\frac{A(r)}{B(r)}(1-g(r)B(r))}dTdr+g(r)dr^2+ r^2 d S^2\,.
\label{p}
\end{equation}
 Thus, for BH with $A(r)=B(r)=1-\frac{2m}{r}+Z(r)$,  $Z(r)>0$, which has restriction with the usual Painlev\`e gauge, one may make the choice $G(r)=\frac{1}{1+Z(r)}$ and
 the (new) metric results to be
 \begin{equation}
ds^2=-\left(1-\frac{2m}{r}+Z(r)\right)dT^2+2\sqrt{\frac{2m}{r}\frac{1}{(1+Z(r))})}dTdr+
\frac{dr^2}{1+Z(r)}+ r^2 d S^2\,,
\label{ppp}
\end{equation}
and it is well defined for any value of $r$.
 Is is easy to show that all the invariant quantities do not depend on the positive function $g(r)$.


 As an application of the generalized  Painlev\`e gauge, we review the Hawking radiation  for static BHs and WHs,  making  use of so called  tunneling method \cite{PK}  in its covariant variant of Hamilton-Jacobi (HJ) tunneling method \cite{ANSZ}.
 
We begin with mass-less test particle action,
\begin{equation}
I=\int_\gamma \partial_\mu I dx^\mu\,,\label{action}
\end{equation}
where $\gamma$ represents a path crossing the horizon.  The action satisfies the HJ relativistic equation,
\begin{equation}
g^{\mu\nu}\partial_\mu I\partial_\nu I=0\,. \label{var}
\end{equation}
Thus, the radial trajectory of a 
mass-less particle is given by,
\begin{equation}
\gamma^{a b} \partial_aI  \partial_b I=0\,.\label{nulltrajectory}
\end{equation}
The relevant two dimensional normal metric in the generalized Pailev\`e gauge (\ref{p})  reads
\begin{equation}
d\gamma^2=-A(r)dT^2-2Q(r)dTdr+g(r)dr^2\,, \quad Q(r)=\sqrt{\frac{A(r)}{B(r)}-g(r)A(r)}   \,.
\end{equation}
Introducing the particle energy $E=-\partial_TI$, one has
\begin{equation}
\partial_r I=-\frac{E}{A(r) }(Q(r)+\sqrt{Q^2(r)+g(r)A(r)})\,.
\end{equation}
Thus,
\begin{equation}
I=- E\int_\gamma \frac{1}{A(r) }(Q(r)+\sqrt{(Q^2(r)+g(r)A(r))}\,  dr\,, 
\end{equation}
with  the integration variable $r$  crossing   the horizon.
We remember that the trapping horizon is located at $r=r_H$ with
$B(r_H)=0\,, B'(r_H)>0$.
First, if we are dealing with a regular WH,  only $B(r_H)=0$,  $A(r_H)\neq 0$ and  $Q(r)$ has a integrable singularity at $r=r_H$. Thus the action is finite and real. On the other side,
in the BH case, also $A(r_H)$ is vanishing at the horizon, and one has
\begin{equation}
B(r)=B'(r_H)(r-r_H)+...\,,\quad\,A(r)=A'(_H)(r-r_H)+...\,.
\label{nh}
\end{equation}
As a consequence, $Q(r_H)=\sqrt{A'(r_H)/B'(r_H)}$. One may split the integration over $r$, and write
\begin{equation}
 I=- E\int_\gamma \frac{1}{A'(r_H)(r-r_H+i\varepsilon) }(Q(r)+\sqrt{(Q^2(r)+g(r)A(r))})\,  dr+I_1\,,
\end{equation}
where $I_1$ is a finite  real  contribution and in the first integral the  horizon divergence is present and  it has been  cured by deforming in a suitable way the  integration in $r$. As a result,
an imaginary part of the action   appears as,
\begin{equation}
\mbox{Im} I= \frac{2\pi E}{\sqrt{A'(r_H)B'(r_H)}}\,. 
\end{equation}
Since the tunneling probability  is given by
\begin{equation}
\Gamma=e^{-2 Im[ I]}\,,
\end{equation}
one derives the  Hawking radiation formula,  
\begin{equation}
\Gamma=e^{- \frac{4 \pi E}{\sqrt{A'(r_H)B'(r_H)}}}\,,
\end{equation}
with Hawking temperature,
\begin{equation}
T_H =\frac{\sqrt{A'(r_H)B'(r_H)}}{4 \pi}\,. 
\end{equation}
 It should be noted that the function $g(r)$ does not enter in the final result.


\end{document}